\documentclass[12pt,preprint]{aastex}

\slugcomment{paper draft to ApJL (ver. 06/09/04)}

\shorttitle{Bipolar Flow toward IRAS 19312+1950}
\shortauthors{Nakashima \& Deguchi}

\begin{document}


\title{Detection of Bipolar Flow toward an Unusual SiO Maser Source IRAS 19312+1950}


\author{Jun-ichi Nakashima\altaffilmark{1} and Shuji Deguchi\altaffilmark{2}}

\altaffiltext{1}{Department of Astronomy, University of Illinois at Urbana-Champaign,
1002 W. Green St., Urbana, IL 61801; junichi@astro.uiuc.edu}

\altaffiltext{2}{Nobeyama Radio Observatory, National Astronomical Observatory, Minamimaki, Minamisaku, Nagano 384-1305, JAPAN; deguchi@nro.nao.ac.jp}


\begin{abstract}
We report a result of an interferometric observation toward an SiO maser source, IRAS 19312+1950, in the HCO$^{+}$ $J=1$--0 line with the Berkeley-Illinois-Maryland-Association (BIMA) Millimeter Array. In the spatially integrated spectrum of HCO$^{+}$, two kinematic components were seen: a strong line with a narrow width ($\sim$ 2 km s$^{-1}$, narrow component), and a weak symmetric line with a broad width ($\sim 60$ km s$^{-1}$, broad component). The line profile of HCO$^{+}$ are reminiscent of that of CO. In the integrated intensity map, we found a pronounced bipolar shape consisting of lower-velocity parts of the broad component. The higher-velocity part of the broad component originated in a relatively small region ($\lesssim 3''$). The spatial structure of the narrow component clearly correlated with near-infrared structure. The position-velocity diagrams indicated presence of a bipolar outflow with an expansion velocity of about 10 km s$^{-1}$. On the basis of the present results, we suggest that the nature of the bipolar flow seen in IRAS 19312+1950 is explained by hydrodynamical interaction between an AGB wind and ambient material with axial symmetric structure. 
\end{abstract}




\section{Introduction}

IRAS 19312+1950 is an SiO maser source discovered by \citet{nak00} in their SiO maser surveys toward selected IRAS sources in the Galactic plane \citep[][]{nak03a, nak03b}. \citet{nak00} first suggested that IRAS 19312+1950 is an AGB/post-AGB star evolved from a relatively massive progenitor on the basis of its prominent bipolar shape seen in near-infrared images and the low color temperature of the dust envelope. In fact, SiO maser sources are usually identified as a late-type star with active mass-loss. Subsequently, they have found two remarkable characteristics of this object: (1) two kinematic components in molecular line profiles \citep[strong line with a narrow width of $\lesssim$ 2 km s$^{-1}$; weak-symmetric line with a broad width of $\gtrsim$ 60 km s$^{-1}$,][]{nak04} ; (2) a rich set of molecular species, \citep[][]{nak04,deg03}.  In addition, recent near-infrared observations (Murakawa 2004, in private communication) suggested that there is a ring-like structure with a size of about 10$''$ around the central star. Some of the characteristics found in IRAS 19312+1950 are often seen in dark clouds (or YSOs). However, secure examples of YSOs emitting SiO masers are quite limited; actually, only three: Ori IRc 2, W51 IRs 2 and Sgr B2 MD5 \citep[cf.,][]{has86}. These YSOs with SiO masers lie in the extreme star forming regions (in giant molecular clouds), which are clearly identified as largely extended nebulae in infrared images, whereas no clear star forming activity is seen around IRAS 19312+1950 in infrared images. Thus the evolutionary status of IRAS 19312+1950 is not yet definitely known.

To investigate the nature of this highly unusual SiO maser source, we are currently conducting high-spatial-resolution interferometric observations using the BIMA array in several molecular rotational lines detected by our single-dish observations \citep{nak00, nak04, deg03}. In this paper, we report the first result of the interferometeric observations in the HCO$^{+}$ $J=1$--0 line, and the detection of a bipolar flow, which might be interpreted by interaction between an AGB wind and ambient material.


\section{Observation and Results}

Interferometric observations of IRAS 19312+1950 were made with the Berkeley-Illinois-Maryland-Association (BIMA) Millimeter Array from December 2003 to January 2004. The instrument is described in detail by \citet{wel96}. We observed the HCO$^{+}$ $J=1$--0 line at 89.188526 GHz with the BIMA array consisting of 10 elements in one configuration (B-array). The observations were interleaved every 25 minutes with the nearby point source, 1925+211, to track the phase variations over time. The absolute flux calibration was determined from observations of Uranus, and is accurate to within 20\%. The final map has an accumulated on source observing time of about 15 hrs. Typical single sideband system temperatures ranged from 200 K to 300 K. The velocity coverage was 380 km s$^{-1}$, using three different correlator windows with a bandwidth of 50 MHz each. The velocity resolution was 1.3 km s$^{-1}$. The phase center of the map was 19$^h$33$^m$24.4$^s$, 19$^{\circ}$56$'$54.8$''$ (J2000) corresponding to the IRAS position of this object. Data reduction was performed with the MIRIAD software package \citep{sau95}. Standard data reduction, calibration, imaging, and deconvolution procedures were followed. Robust weighting of the visibility data gave 3.7$''$ $\times$ 2.5$''$ CLEAN beam with a position angle of 23.0$^{\circ}$. The r.m.s. noise per 1.0 km s$^{-1}$ is 3.8$\times$10$^{-2}$ Jy beam$^{-1}$.

Figure 1 shows the spectrum of the HCO$^{+}$ $J=1$--0 line. A remarkable feature seen in the spectrum is a strong line with a narrow width ($\sim$ 2 km s$^{-1}$) peaked at $V_{lsr} \sim 38$ km s$^{-1}$. A weak symmetric line with a broad width ($\sim$ 60 km s$^{-1}$) is also seen in the velocity range of $V_{lsr} \sim -10$ --- 70 km s$^{-1}$. For convenience, we call the former the "narrow component", and the latter the "broad component" in this paper. The narrow component is roughly centered on the broad component. Absorption features are seen beside the narrow component, suggesting that cold gas components lie in the foreground. The line profile of HCO$^{+}$ is reminiscent of that of the CO $J=1$--0 (and $J=2$--1) line \citep[cf.,][]{nak04}. Similar line profiles including the two kinematic components (narrow and broad components) have been reported in some AGB stars \citep[ex., EP Aqr, X Her and RV Boo,][]{kah96, ker99, kna98}. \citet{kah96} reported on a bipolar flow consisting of the broad component seen in X Her, and \citet{kna98} suggested a scenario where episodic mass-loss plays a role to explain the kinematic components seen in these AGB stars. Figure 2 shows velocity integrated intensity maps in three different velocity ranges corresponding to the narrow component (broken contour) and the blue- and red-shifted wings of the broad component (thin and thick solid contours, respectively). The map is superimposed on a combined near-infrared image ($J$, $H$, and $K$-bands) taken by the CIAO camera on the SUBARU telescope (courtesy of Koji Murakawa, Motohide Tamura and the CIAO group in the National Astronomical Observatory in Japan). The velocity ranges used for computing integrated intensities are indicated in Figure 1 and the caption of Figure 2. The source is clearly resolved by the synthesized beam (shown at the lower-right in Figure 2). Though we observed with only one configuration, most part of the flux emitted from the source is thought to be detected according to the upper limit of the source size ($\sim$ 30$''$) determined by our single-dish observation \cite[][]{nak04}. As the most pronounced feature in the map, we can see a bipolar shape consisting of the thin and thick contours. The apparent axis of the bipolar shape is close to the NNE --- SSW direction. If the bipolar shape originates in a bipolar outflow, the lower side (southern component) should be closer to us, and the upper side (northern component) should be further from us. Interestingly, the structure of the blue-shifted wing (thin contour) of the broad component shows two extended tails: the strong tail to the southeast, and the weak tail to the east exhibiting a "lying-Y" shape. The structure of the red-shifted wing (thick contours) extends only to the northeast. On the contrary, the structure of the narrow component (broken contour) is somewhat complicated. Three intensity peaks are seen in the broken contour at the points of 3$''$, 4$''$ and 8$''$ from the map center to the northeast, northwest and east, respectively. Although the positions of these peaks of the narrow component do not correspond to bright regions seen in the near-infrared image (background of Figure 2), it looks as though the global structure of the narrow component clearly correlates with the near-infrared structure. From the apparent size of the HCO$^{+}$ structure and the assumption of the density, we can crudely estimate the mass of the source. If we assume that the source is a homogeneous sphere with a diameter of 10$''$, the mass is estimated to be 4--31 M$_{\odot}$, where we used the distance (2.5 -- 5.5 kpc) estimated by \cite{nak04} and the assumed density of 10$^4$ cm$^{-3}$ traced by the HCO$^{+}$ line.

To investigate the kinematic structure, we made position-velocity ($p$--$v$) diagrams at various cuts. Figure 3 shows the $p$--$v$ diagrams at selected cuts. In the top four diagrams in Figure 3, velocity channels are averaged over 2 km s$^{-1}$ intervals. The cuts used for the diagrams are indicated in Figure 2 as the broken arrows. The directions of the arrows in Figure 2 represent the positive direction of the offset axes in Figure 3. The origins of the cuts A, B, C and D are taken at the phase center. The origin of the cut E is indicated by the filled dot in Figure 2. The cut "A" corresponds to the apparent axis of the bipolar shape. The cut "B" represents the perpendicular cut to the cut A. The cut "C" corresponds to the major axis of the ring-like structure seen in near-infrared images. The cut "D" represents the perpendicular cut to the cut C. The cut "E" is mentioned later. Although the global structure seen in Figure 3 is highly clumpy, the spatial size (vertical size of structure) of the source tends to be smaller in higher-velocity regions.  Intensity peaks lie nearly on the origins in vertical axes in velocity ranges, $V_{lsr} \lesssim 20$ km s$^{-1}$ and $V_{lsr} \gtrsim 50$ km s$^{-1}$. In addition, maximum velocity widths (represented by the maxim widths of the lowest contours in the horizontal coordinates) do not exhibit any noticeable variations with respect to the position angle. These facts mean that the higher velocity part ($|V_{lsr} - V_{sys}| \gtrsim 15$ km s$^{-1}$; here, we assume that the system velocity is $V_{sys} \sim 37$ km s$^{-1}$) of the broad component is not resolved by our synthesized beam, and should originate in a relatively small region ($\lesssim 3''$). On the other hand, variations of structure are found in lower velocity ranges ($|V_{lsr} - V_{sys}| \lesssim 15$ km s$^{-1}$) of the broad component especially in the blue-shifted side (20 km s$^{-1}$ $\lesssim$ $V_{lsr} \lesssim$ 35 km s$^{-1}$). In fact, intensity peaks at $V_{lsr} \sim 25 - 27$ km s$^{-1}$ clearly shift to the negative direction in the offset axes. These peaks correspond to the southern and northern tails of the blue-shifted wing (cf., thin contour in Figure 2). The 27 km s$^{-1}$ peaks seen in the cuts A, C and D correspond to the southern tail, and the 25 km s$^{-1}$ peak seen in the cut B corresponds to the northern tail. On the contrary, in the red-shifted side, no clear variations of structure are seen except for an extension to the positive direction in the offset axis in the cut A (and possibly in C) at $V_{lsr} \sim 47$ km s$^{-1}$. We also made $p$--$v$ diagrams without channel binding to check the motion of the narrow component in several cuts. Finally, we found the most pronounced variation of structure in the cut "E" corresponding to the tail growing to the southeast from the intensity peak at (DEC, RA)$_{\rm offset} =$(1.1$''$, $-$2.5$''$). The $p$--$v$ diagram in the cut E is shown in the bottom panel in Figure 3, and there is clear variation of structure as a function of radial velocity. The peak has another tail extended to the southwest. The southwest tail also exhibits a systematic variation of structure in a $p$--$v$ diagram, though the diagram is not shown in this paper. In any other cuts, no clear variation of velocity structure is found in the narrow component. We are likely to see complicated variation of the narrow component in the top four panels in Figure 3. These structures are explained well by a superposition of the two flows mentioned above.


\section{Discussion}
The bipolar flow seen in IRAS 19312+1950 consists of the lower-velocity part of the broad component, whereas the higher-velocity part originates in a small region with a size of less than about 3$''$. Usually, the expansion velocity of bipolar flows seen in late-type stars (especially in post-AGB stars) tends to increase with a distance from the central star \citep[frequently called "Hubble type flow" in this area, cf.,][]{bal02}, the nature of the bipolar flow seen in IRAS 19312+1950 is in disagreement with this point. In our opinion, the characteristics of the bipolar flow are explained by hydrodynamical interaction between a spherical outflow lying at the center and ambient material with axial symmetric structure if the spherical outflow expelled from the central star is distorted to a bipolar shape by the interaction. In such a case, the slow expanding velocity of the bipolar flow seems to be somewhat strange, but the slow velocity would be explained by a projection effect if the flow has an inclination angle of 20$^{\circ}$ (where we assume the velocity of the bipolar flow equals to a half of the maximum width of the broad component, i.e. $\sim$ 30 km s$^{-1}$). Interestingly, in Figure 2, the bipolar shape precisely escapes from dense regions of the narrow component, and the red-shifted component of the bipolar flow lies in the interspace between two remarkable intensity peaks of the narrow component. In addition, the absorption feature beside the narrow component seen in Figure 1 might be explained as the near side of the ambient material (with axial symmetric structure). 

If the higher-velocity part of the broad component is a spherical outflow from the central star, the flow should be identified as an AGB outflow, because the spherical outflow is the typical nature of AGB envelopes, and also because the expanding velocity of $V_{exp} \sim 30$ km s$^{-1}$ is very reasonable as that seen in the super-wind phase at the late-AGB stage. Our recent CO data taken by the BIMA array also support the spherical property of the inner part of the nebulosity. However, a problem is the somewhat strong HCO$^{+}$ intensity (especially in the outermost parts of the envelope) if the central star is an O-rich AGB star, because the HCO$^{+}$ line is usually weak or null in AGB envelopes except a few cases \citep[][]{ngu88,deg90,cox92}. From the viewpoints of theory, the HCO$^{+}$ molecule can be produced by photo-chemical process in an O-rich AGB envelope up to the radius $\sim 2 \times 10^{17}$ cm \citep[e.g.,][; Mamon et al. took account of ionization by galactic and stellar UV photons, and high-energy particles in their calculation.]{mam87}. However, we still need additional chemical process to explain the effective formation of HCO$^{+}$ in the outermost parts of the envelope, because it is most likely that the outermost part of HCO$^{+}$ emission is further than 10$^{17}$ cm from the central star if the central star is an AGB star \citep[estimated distance: 2.5--5.1 kpc, see][]{nak04}. An immediate possibility to explain the highly extended HCO$^{+}$ structure would be "dissociative shock". Actually, most AGB envelopes exhibiting the HCO$^{+}$ emission show signs of shocks \citep[cf.,][]{cox92}. The HCO$^{+}$ emission in OH 231.8+4.2, which is a well-known example of an SiO maser source with a bipolar shape, is also explained by shocks \citep{san00}. The shock ionization also supports our idea to explain the kinematic structure of the bipolar flow, because shock fronts are likely to be produced between an AGB wind and ambient material.

Remaining problems in the interpretation as an AGB star are excess of mass and mass-loss rate. Although we estimated mass to be 4--31 M$_{\odot}$ in a previous section, the mass will be over 8 M$_{\odot}$ if the distance exceeds 3 kpc. In such a case, the nebulosity cannot be explained as material expelled from a central AGB star. If we assume a spherical flow and density of 10$^{4}$ cc$^{-1}$ at 10$^{17}$ cm from the central star, the mass-loss rate is estimated to be about 10$^{-4}$ M$_{\odot}$ yr$^{-1}$ in terms of the expanding velocity of the broad component ($\sim$ 30 km s$^{-1}$). This mass-loss rate is slightly larger than a typical value of AGB stars ($\sim 10^{-5}$ M$_{\odot}$ yr$^{-1}$). The keys to check the AGB possibility would be near-star kinematics and isotope ratio (ex., $^{13}$C/$^{12}$C), because AGB stars and YSOs are known to be have significantly different kinematics especially in a near-star region, and because $^{13}$C should be enhanced in a circumstellar envelope if the central star lies at the late-AGB stage.


\section{Summary}
In this paper, we reported a result of an interferometeric observation of an SiO maser source, IRAS 19312+1950, in the HCO$^{+}$ $J=1$--0 line. We found two kinematical components in the spectrum including a strong-narrow line (narrow component) and a weak-broad line (broad component). Velocity integrated intensity maps showed a pronounced bipolar shape consisting of lower-velocity parts of the broad component. The higher-velocity parts of the broad component were not resolved by our synthesized beam, and should originate in a relative small region with the size of less than $\sim$ 3$''$. The spatial structure of the narrow component clearly correlated with the near-infrared structure. In $p$--$v$ diagrams, a systematic variation of structure was found in the lower-velocity part of the broad component, indicating a presence of a bipolar outflow. We suggest that the characteristics of the bipolar flow seen in IRAS 19312+1950 are explained by interaction between an AGB wind and ambient material with axial symmetric structure.


\acknowledgments

The authors are grateful to Kee-Tae Kim, James R. Forster, Hiroshi Imai, and Hideyuki Izumiura for stimulating discussion, and to Anthony Remijan for help with the BIMA observations and data reduction. We would like to also thank Douglas Friedel for help in correction of the English, and the anonymous referee for useful comments. This research has been supported by the Laboratory for Astronomical Imaging at the University of Illinois and NSF AST 0228953, and has made use of the SIMBAD and ADS databases.

\begin{figure}
\epsscale{.60}
\plotone{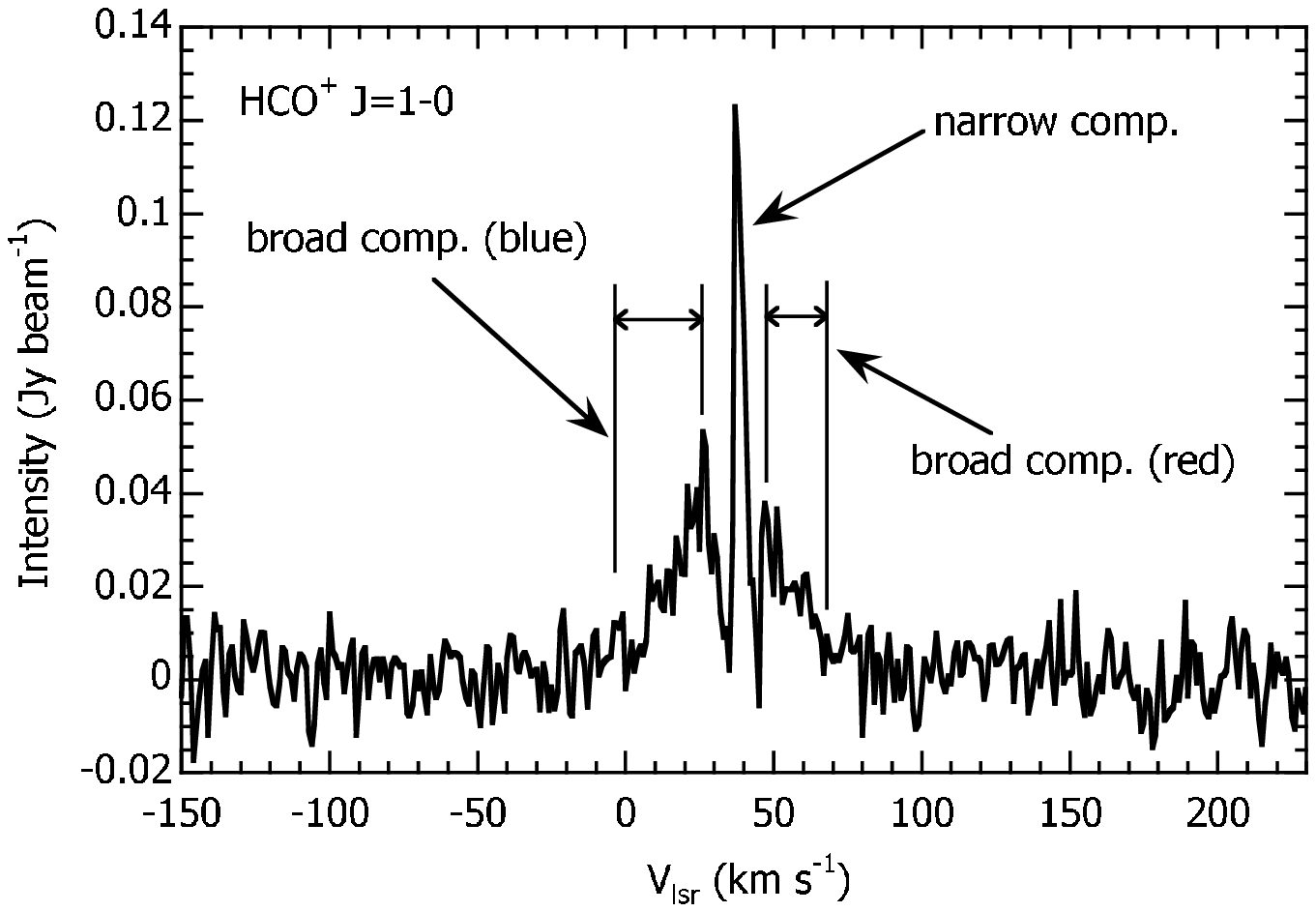}
\caption{Spatially integrated spectrum of IRAS 19312+1950 in the HCO$^{+}$ $J=1$--0 line. The integrated area is a circle with a diameter of 15$''$. The vertical solid lines represent the velocity ranges of the blue- and red-shifted wings of the broad component used for Figure 2. \label{fig1}}
\end{figure}

\clearpage

\begin{figure}
\epsscale{.60}
\plotone{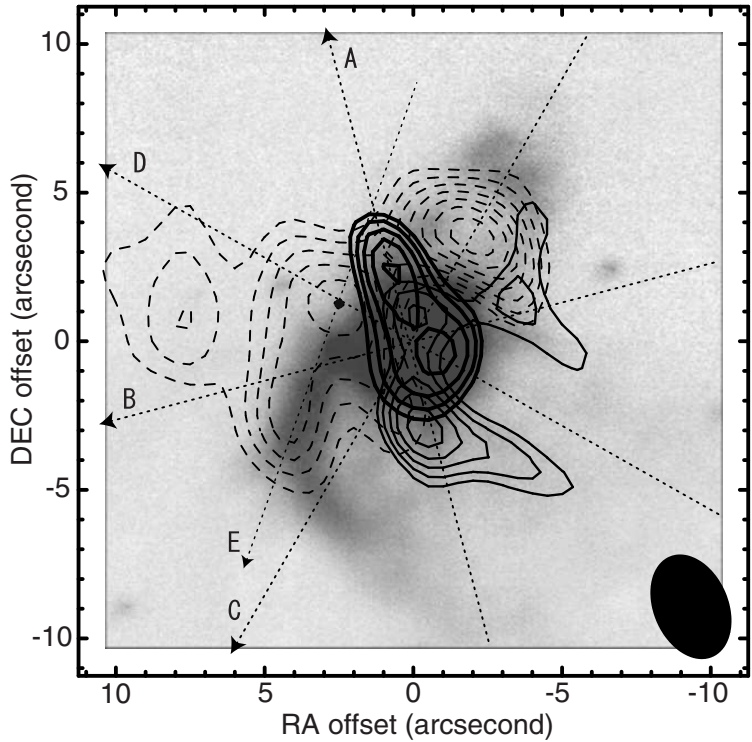}
\caption{HCO$^{+}$ integrated intensity map (contours) superimposed on a near-infrared composite (J, H and K bands) image (grayscale) taken by the SUBARU telescope. The ranges of velocity integration for the thick, thin and broken contour maps are 0 -- 26 km s$^{-1}$, 47 -- 60 km s$^{-1}$ and 37 -- 38 km s$^{-1}$, respectively. The contour levels start from a 5 $\sigma$ level , and increase by every 0.5 $\sigma$ for the broad component and by every 1 $\sigma$ for the narrow component. The 1 $\sigma$ levels of the thin, thick and broken contour maps are $7.3 \times 10^{-3}$ Jy beam$^{-1}$, $1.0 \times 10^{-2}$ Jy beam$^{-1}$ and $2.7 \times 10^{-2}$ Jy beam$^{-1}$, respectively. The synthesized beam is indicated in the lower right corner. The broken arrows (A, B, C, D and E) denote "cuts" used for $p$--$v$ diagrams in Figure 3. The filled dot represents the origin of the cut E. \label{fig2}}
\end{figure}

\clearpage

\begin{figure}
\epsscale{.30}
\plotone{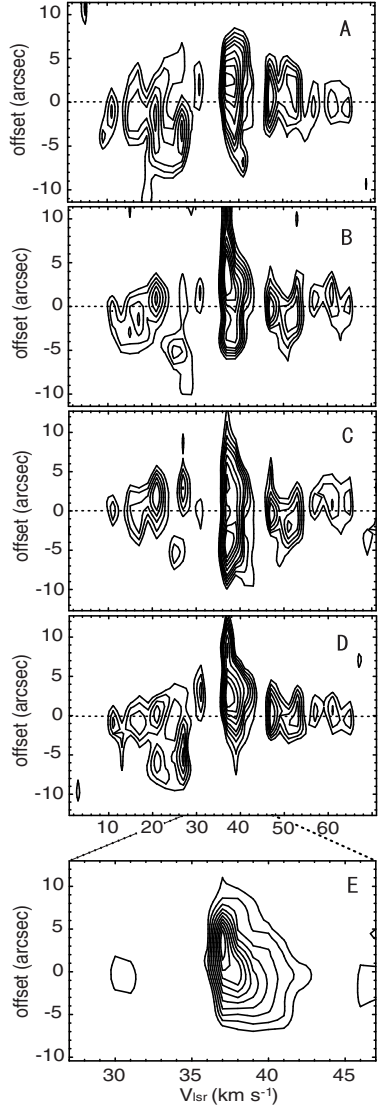}
\caption{Position-velocity diagrams along the cuts indicated in Figure 2. The velocity channels are averaged over 2 km s$^{-1}$ intervals except for the bottom panel. In the top four diagrams, the contours are drawn from 40 mJy beam$^{-1}$ with increments of 10 mJy beam$^{-1}$ between 40 and 80 mJy beam$^{-1}$, and 20 mJy beam$^{-1}$ between 80 and 160 mJy beam$^{-1}$. The bottom panel is the $p$--$v$ diagram along the cut "E" without velocity channel binding. In the bottom panel, the contours start from 60 mJy beam$^{-1}$ with increments of 20 mJy beam$^{-1}$. \label{fig3}}
\end{figure}


\begin{thebibliography}{}
\bibitem[Balick \& Frank(2002)]{bal02} Balick, B. \& Frank, A.  \araa, 40, 439
\bibitem[Cox et al.(1992)]{cox92} Cox, P., Omont, A., Huggins, P., Bachiller, R. \& Forveille, T.  1992, \aap, 266, 420
\bibitem[Deguchi et al.(1990)]{deg90} Deguchi, S., Izumiura, H., Kaifu, N., Mao, X., Nguyen-Q-Rieu, \& Ukita, N. 1990, ApJ, 351, 522
\bibitem[Deguchi \& Nakashima(2003)]{deg03} Deguchi, S. \& Nakashima, J.  2003, in Proc. IAU Symp. 209, Planetary Nebulae: Their Evolution and Role in the Universe (San Franscisco: ASP), 259
\bibitem[Hasegawa et al.(1986)]{has86} Hasegawa, T. et al.  1986, Masers/Molec. \& Mass Outflows / Star Form. Regions, p.275
\bibitem[Kahane \& Jura(1996)]{kah96} Kahane, C. \& Jura, M.  1996, \aap, 310, 952
\bibitem[Kerschbaum \& Olofsson(1999)]{ker99} Kerschbaum, F. \& Oloffson, H.  1999, \aaps, 138, 299
\bibitem[Knapp et al.(1998)]{kna98} Knapp, G. R., Young, K., Lee, E. \& Jorrissen, A.  1998, \apjs, 117, 209
\bibitem[Mamon, et al.(1987)]{mam87} Mamon, G. A., Glassgold, A. E., \& Omont, A. 1987, ApJ, 323, 306
\bibitem[Nakashima \& Deguchi(2000)]{nak00} Nakashima, J. \& Deguchi, S.  2000, \pasj, 52, L43
\bibitem[Nakashima \& Deguchi(2003a)]{nak03a} Nakashima, J. \& Deguchi, S.  2003a, \pasj, 55, 203
\bibitem[Nakashima \& Deguchi(2003b)]{nak03b} Nakashima, J. \& Deguchi, S.  2003b, \pasj, 55, 229
\bibitem[Nakashima et al.(2004)]{nak04} Nakashima, J., Deguchi, S. \& Kuno, N.  2004, \pasj, 56, 193
\bibitem[Nguyen-Q-Rieu et al.(1988)]{ngu88} Nguyen-Q-Rieu, Deguchi, S., Izumiura, H., Kaifu, N., Ohishi, M., Suzuki, H., Ukita, N. 1988, ApJ, 330, 374
\bibitem[S\`anchez Contreras et al.(2000)]{san00} S\`anchez Contreras, C., Bujarrabal, V., Neri, R. \& Alcolea, J.  2000, \aap, 357, 651
\bibitem[Sault et al.(1995)]{sau95} Sault, R. J., Teuben, P. J. \& Wright, M. C. H.  1995, Astronomical Data Analysis Software and Systems IV, ASP Conference Series, Vol. 77, eds., Shaw, R. A., Payne, H. E., \& Hayes, J. J. E., p.433
\bibitem[Welch et al.(1996)]{wel96} Welch, W. J., et al.  1996, \pasp, 108, 93
\end{thebibliography}
\end{document}